# Magnetism and structure of Li$_x$CoO$_2$ and comparison to Na$_x$CoO$_2$

J.T. Hertz[1], Q. Huang[2], T. McQueen[1], T. Klimczuk[3,4], J.W.G. Bos[5], L. Viciu[1], and R.J. Cava[1]

[1]Department of Chemistry, Princeton University, Princeton NJ 08544
[2]Center for Neutron Research, NIST, Gaithersburg MD 20899
[3]Divison of Thermal Physics, Los Alamos National Laboratory, Los Alamos NM 87545
[4]Faculty of Applied Physics and Mathematics, Gdansk University of Technology,
Narutowicza 11/12, 80-952 Gdansk, Poland,
[5] School of Chemistry and Centre for Science at Extreme Conditions, University of Edinburgh, Edinburgh, EH9 3JJ, United Kingdom

**Abstract**

The magnetic properties and structure of Li$_x$CoO$_2$ for $0.5<x<1.0$ are reported. Co$^{4+}$ is found to be high-spin in Li$_x$CoO$_2$ for $0.94 \leq x \leq 1.00$ and low-spin for $0.50 \leq x \leq 0.78$. Weak antiferromagnetic coupling is observed, increasing in strength as more Co$^{4+}$ is introduced. At $x \approx 0.65$, the temperature-independent contribution to the magnetic susceptibility and the electronic contribution to the specific heat are largest. Neutron diffraction analysis reveals that the lithium oxide layer expands perpendicular to the basal plane and the Li ions displace from their ideal octahedral sites with decreasing x. A comparison of the structures of the Na$_x$CoO$_2$ and Li$_x$CoO$_2$ systems reveals that the CoO$_2$ layer changes substantially with alkali content in the former but is relatively rigid in the latter. Further, the CoO$_6$ octahedra in Li$_x$CoO$_2$ are less distorted than those in Na$_x$CoO$_2$. We postulate that these structural differences strongly influence the physical properties in the two systems

## Introduction

The Na$_x$CoO$_2$ system has recently been under intense study because it exhibits remarkable electronic properties such as anomalously high thermopower [1], superconductivity when intercalated with water [2], and an unexpected metal-insulator transition [3]. These observations have motivated investigation into the relationships between structure and properties in phases based on NaCoO$_2$ [see e.g. 4-8]. The structures of four different two- and three-layer Na$_x$CoO$_2$ phases, which are stable in various compositional ranges, have been determined. As the sodium content is varied, the coordination of the sodium ions, the layer alignment, and the unit cell parameters change considerably. These general structural trends have been correlated to trends in the electronic and magnetic behavior via structural distortions of the CoO$_2$ layer [4,10]. The Li$_x$CoO$_2$ system has been studied extensively in regards to its application in rechargeable batteries [see e.g. 11-13], but significantly less attention has been paid to its structural and magnetic properties.

Three forms of LiCoO$_2$ have been reported: thermodynamically stable O3-LiCoO$_2$ (or HT-LiCoO$_2$), a three-layer rhombohedral compound; O2-LiCoO$_2$, a metastable two-layer hexagonal compound synthesized through exchange reactions with P2-Na$_{0.7}$CoO$_2$ [14-16]; and spinel-like LiCoO$_2$ (or LT-LiCoO$_2$), synthesized by soft chemistry methods [17,18]. In the present investigation, we limit our study to thermodynamically stable three-layer HT-LiCoO$_2$, which is structurally analogous to three layer NaCoO$_2$. Lithium deintercalation of LiCoO$_2$ has been accomplished through the use of a wide variety of oxidizing reagents, such as I$_2$, Br$_2$, Cl$_2$, H$_2$SO$_4$, HCl, NO$_2$BF$_4$ and NO$_2$PF$_6$ [19-23]. Electrochemical extraction of lithium is also common [24-29]. When lithium ions are removed to form Li$_x$CoO$_2$, 1-x cobalt atoms are formally oxidized to Co$^{4+}$.

LiCoO$_2$ has the layered α-NaFeO$_2$-type structure, consisting of sheets of Co$^{3+}$-based edge-sharing CoO$_6$ octahedra separated from one another by sheets of Li$^+$. The Li$^+$ occupy octahedral sites between the oxygens of adjacent cobalt oxide layers. The CoO$_2$ layers stack in an ABCABC fashion, and one unit cell of LiCoO$_2$ contains three layers in rhombohedral symmetry, in space group *R-3m* [30-32]. Some magnetic characterization of Li$_x$CoO$_2$ has previously been reported [21, 33-37] but comprehensive structure-property correlations have not yet been established. The spin state of Co$^{3+}$ and Co$^{4+}$ ions in oxides is a matter of active study, but such ions are generally found to be in their low spin $d^6$ and $d^5$ configurations in compounds like Li$_x$CoO$_2$ and Na$_x$CoO$_2$ [1,38-43].

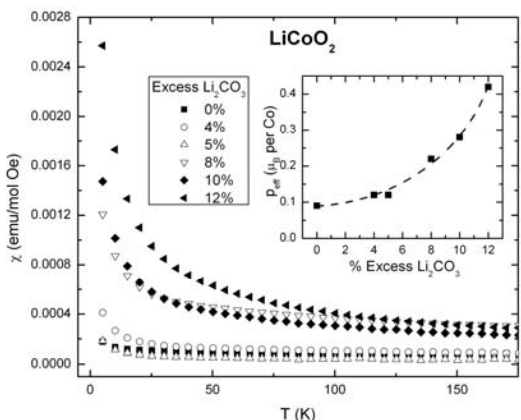

**Fig 1.** Temperature dependent magnetic susceptibility of $LiCoO_2$ for samples made with various amounts of excess $Li_2CO_3$. Inset: The effective magnetic moments per Co of the nominal $LiCoO_2$ samples plotted vs. the amount of excess $Li_2CO_3$ used in the synthesis.

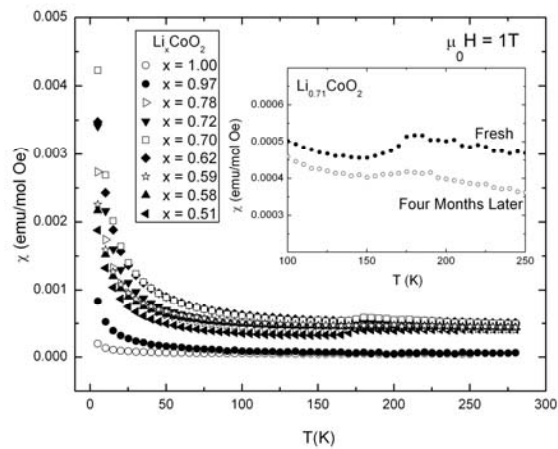

**Fig. 2**: Temperature dependent magnetic susceptibilities for $Li_xCoO_2$ Inset: Comparison of the magnetic susceptibility of $Li_{0.72}CoO_2$ as freshly made and after storage for four months.

The present study characterizes $Li_xCoO_2$ in the range $0.5 \leq x \leq 1.0$, the compositions accessible to soft chemistry methods [25,44]. Previous studies have shown that $Li_xCoO_2$ maintains a three-layered hexagonal structure for $0.54 \leq x \leq 1.00$, with a mixture of two hexagonal phases in the range $0.78 \leq x \leq 0.94$ [25-27,45]. A monoclinic distortion has been observed in the region $0.46 < x < 0.54$. The structure of this monoclinically distorted compound has been studied by electron diffraction [28], and the structures of some other members of the $Li_xCoO_2$ system have been studied by X-ray diffraction [32]. Here we present the magnetic and structural characterization of a uniform series of $Li_xCoO_2$ samples, comparing those results to the same characteristics in $Na_xCoO_2$, with the aim of establishing broad structure-property correlations in electronic systems based on hexagonal $CoO_2$ planes.

**Experimental**

The $LiCoO_2$ used in the present study was synthesized by previously established methods [e.g. refs. 19-21, 46-47]. Stoichiometric amounts of $Li_2CO_3$ (Alfa Aesar, 99%) and $Co_3O_4$ (Alfa Aesar, 99.7%) were combined and mixed thoroughly. The mixture was placed in a covered alumina crucible, heated to 900°C, and held for 24 hours under a slow flow of oxygen. The samples were then furnace cooled, thoroughly ground, and heated at 900°C for another 24 hours under flowing oxygen. The identity and purity of the $LiCoO_2$ was confirmed by powder X-ray diffraction (Bruker D8 focus, Cu Kα radiation, diffracted beam graphite monochromator). Powder X-ray diffraction patterns were also used to characterize the crystallographic cell parameters of the deintercalated samples. Internal silicon standards were employed, and unit cell refinement was accomplished using Topas (Bruker XRD) software.

In some previously reported syntheses of $LiCoO_2$, a small excess of lithium carbonate was used to counterbalance loss of lithium oxide, which vaporizes at high temperatures [19,21,23]. The use of excess $Li_2CO_3$ is important in the present study; because Li excess $Li_{1+x}CoO_2$ has been previously reported [37]. In agreement with that study we find that the magnetic properties near x=1 in $Li_xCoO_2$ are dependent on the initial mixture of starting materials. In order to evaluate the effects of using excess lithium on the synthesis and magnetic properties of $LiCoO_2$, six samples of $LiCoO_2$ were prepared as described above using 0%, 4%, 5%, 8%, 10%, and 12% molar excess $Li_2CO_3$.

For the magnetic susceptibility and specific heat studies, a 10 gram master batch of nominally stoichiometric $LiCoO_2$ was synthesized. For the deintercalation of the starting material, 0.5 g samples of $LiCoO_2$ were reacted with different amounts of liquid $Br_2$ (Alfa Aesar, 99.8%). Between 0.05 mL and 2.5 mL of bromine was added to each 0.5 g sample of $LiCoO_2$ in 10 mL of acetonitrile, and the reaction mixture was stirred for three days [48]. The deintercalated samples were filtered, washed with acetonitrile, and dried under aspiration. All $Li_xCoO_2$ samples with compositions between x=0.5 and x=1.0 were found to be stable in dry air. For determination of the Li content, 4 mg of each $Li_xCoO_2$ sample was dissolved in 2 mL of HCl and diluted with 18 mL of deionized $H_2O$. ICP-AES analysis was performed using a Perkin Elmer Optima 4300 ICP-AES and powder X-ray diffraction was performed [49].

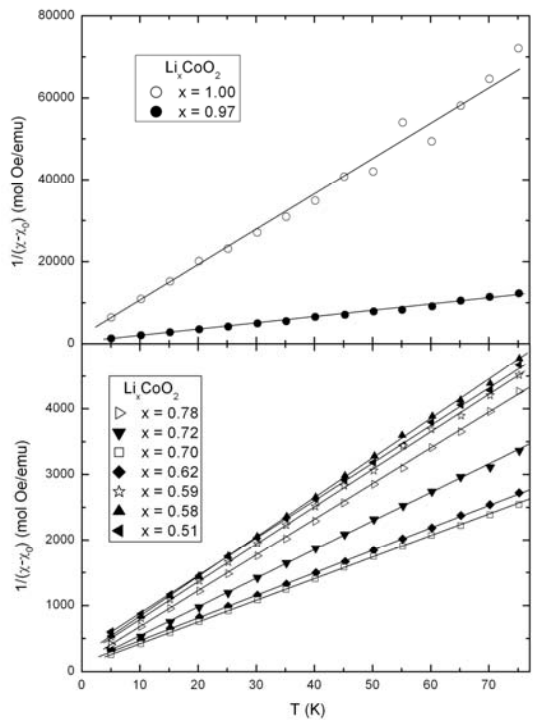

**Fig. 3**: Upper panel: The temperature dependence of $1/(\chi - \chi_0)$ for the two $Li_xCoO_2$ samples above the two-phase region over the range 5 K ≤ T ≤ 75 K. Lower panel: The temperature-dependence of $1/(\chi - \chi_0)$ for the seven $Li_xCoO_2$ samples below the two-phase region over the range 5 K ≤ T ≤ 75 K.

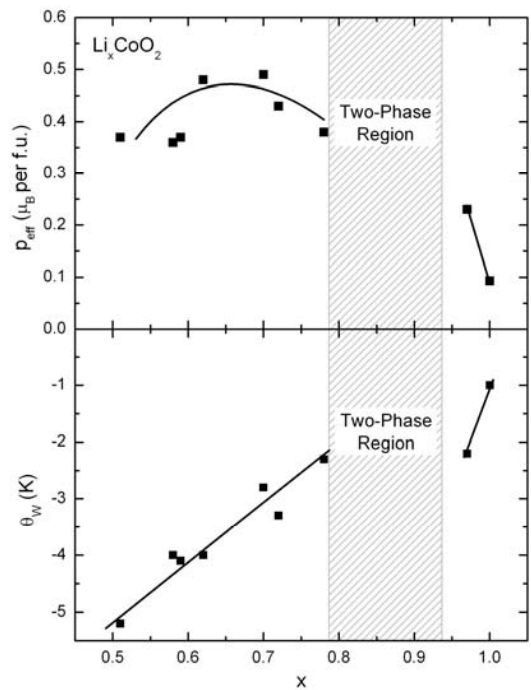

**Fig. 4**: Upper panel: The variation of $p_{eff}$ per formula unit with $x$ for $Li_xCoO_2$. Lower panel: The variation of the Weiss temperature ($\theta_W$) with $x$ for $Li_xCoO_2$.

For neutron diffraction measurements, a second 10 gram master batch of nominally stoichiometric $LiCoO_2$ was synthesized. Five samples of $Li_xCoO_2$ (0.5≤$x$≤1.0) were then made by reacting 2 g aliquots of this $LiCoO_2$ with the $Br_2$ oxidant solutions. One of the samples used in the neutron diffraction study was selected to be within the two-phase region (0.78≤$x$≤0.94) [12,26,50,51] so that the structures of the $Li_xCoO_2$ compounds at the border compositions ($x$ = 0.78 and $x$ = 0.94) could be determined.

Neutron diffraction analysis was performed at the NIST center for neutron research on the BT1 diffractometer. The powder neutron diffraction patterns at 298 K were obtained using a Cu (311) monochromator with a 90° take-off angle, λ = 1.5404(2) Å, and in-pile and diffracted beam collimations of 15' and 20' respectively. Data were collected over the two theta range 3–168° with a step size of 0.05°. The GSAS program suite was used for Rietveld structural refinement [52]. Scattering lengths (in *fm*) employed in the refinement were -1.90, 2.49, and 5.80 for Li, Co, and O, respectively.

The magnetic susceptibilities and specific heats of the samples were measured in a Quantum Design Physical Properties Measurement System (PPMS). Specific heat measurements were performed on six of the $Li_xCoO_2$ samples used for magnetic analysis ($x$ = 1.00, 0.78, 0.70, 0.62, 0.58, and 0.51). About 25 mg of sample powder was mixed with an equal mass of Ag powder, and then pressed into a pellet. The use of silver was necessary because $Li_xCoO_2$ samples are poor thermal conductors. The specific heat of each sample was measured at low temperatures (2 K ≤ T ≤ 15 K), and the contribution of the Ag powder was subtracted.

**Magnetism and stiochiometry in LiCoO₂**

The laboratory powder X-ray diffraction patterns obtained from the samples of $LiCoO_2$ synthesized with excess $Li_2CO_3$ were virtually indistinguishable from the pattern obtained from the sample synthesized with no excess lithium, indicating that the samples possess very similar structures and lattice parameters. Thus, the crystal structure of the product was not strongly affected by the amount of excess lithium present in the reaction mixture. Furthermore, none of the XRD patterns contained detectable impurity peaks, indicating that a lithium-rich impurity phase did not form in significant proportion in any of the samples. Laboratory XRD analysis, therefore, provided little information about how much, if any, excess $Li_2CO_3$ is needed to synthesize stoichiometric $LiCoO_2$. Magnetic measurements, which are more sensitive, were therefore used to distinguish the samples.

The average oxidation state of cobalt is directly related to the amount of lithium in the compound. If lithium is lost to vaporization during synthesis and $Li_xCoO_2$ ($x<1.00$) is formed, then $1-x$ cobalt atoms will be oxidized to $Co^{4+}$. Similarly, if extra lithium is incorporated into a compound of either $Li_{1+x}CoO_2$ or $Li_{1+x}Co_{1-x}O_2$ ($x>0$), then $x$ (2x) cobalt atoms will be reduced (oxidized) to $Co^{2+}$ ($Co^{4+}$) to maintain charge neutrality. Thus, if a nonstoichiometric amount of Li is incorporated into $LiCoO_2$, the sample will show a magnetic moment: only perfectly stoichiometric $LiCoO_2$ will show no local moment because its $Co^{3+}$ is in a low-spin $d^6$ configuration. By measuring the magnetic susceptibilities of the six samples made with varying amounts of $Li_2CO_3$, it was possible to determine which sample was least magnetic and thus closest to stoichiometric $LiCoO_2$.

The results of the magnetic susceptibility measurements on these samples are displayed in Figure 1. As the figure shows, the samples become more magnetic with increasing $Li_2CO_3$. Curie-Weiss fits were performed using $\chi = \chi_0 + C/(T-\theta_{cw})$, where $\chi$ and $\chi_0$ are the measured and temperature independent parts of the susceptibility, C is the Curie constant, T is temperature in Kelvin and $\theta_{cw}$ is the Curie-Weiss temperature. The calculated effective magnetic moments per Co for each of the six samples are displayed in the inset to Figure 1. The use of excess $Li_2CO_3$ results in the presence of a magnetic moment that increases in magnitude as higher lithium-containing starting material is used. The least magnetic sample was the one made without excess $Li_2CO_3$, and, therefore, no excess $Li_2CO_3$ was used in subsequent syntheses of $LiCoO_2$.

**Table 1**: Magnetic and electronic characterization of $Li_xCoO_2$.

| Li Content ($x$) | $\chi_{280}$ (emu/mol$_{Co}$Oe) | Low Temp. $\chi_0$ (emu/mol$_{Co}$Oe) | $\theta_W$ (K) | $p_{eff}$ ($\mu_B$/Co) | $p_{eff}$ ($\mu_B$/Co$^{4+}$) | $\Gamma$ mJ/molK$^2$ |
|---|---|---|---|---|---|---|
| 1.00 (0.985) | $0.44 \times 10^{-4}$ | $0.43 \times 10^{-4}$ | -1.0 | 0.09 | -- (5.9) | 0.4 |
| 0.97 (0.961) | $0.63 \times 10^{-4}$ | $0.33 \times 10^{-4}$ | -2.2 | 0.23 | 7.67 (5.9) | - |
| 0.78 | $4.14 \times 10^{-4}$ | $2.78 \times 10^{-4}$ | -2.3 | 0.38 | 1.72 | 12.4 |
| 0.72 | $4.51 \times 10^{-4}$ | $2.71 \times 10^{-4}$ | -2.3 | 0.43 | 1.54 | - |
| 0.70 | $4.96 \times 10^{-4}$ | $3.13 \times 10^{-4}$ | -2.8 | 0.49 | 1.63 | 18.9 |
| 0.62 | $5.08 \times 10^{-4}$ | $3.52 \times 10^{-4}$ | -4.0 | 0.48 | 1.26 | 19.9 |
| 0.59 | $4.41 \times 10^{-4}$ | $3.52 \times 10^{-4}$ | -4.1 | 0.37 | 0.90 | - |
| 0.58 | $4.37 \times 10^{-4}$ | $3.28 \times 10^{-4}$ | -4.0 | 0.36 | 0.86 | 18.4 |
| 0.51 | $4.11 \times 10^{-4}$ | $1.80 \times 10^{-4}$ | -5.2 | 0.37 | 0.76 | 14.6 |

The origin of the observed local moment in the Li excess samples is open to speculation. It has been previously proposed that incorporation of extra lithium results in $Li_{1+x}CoO_2$ ($x>0$) and the reduction of $x$ $Co^{3+}$ atoms to $Co^{2+}$ [12,17-19]. This is unlikely for two reasons. Firstly, all the octahedral sites in the lithium layer are filled in stoichiometric $LiCoO_2$, so there is no room for more due to Li-Li repulsion. Secondly, the synthesis is performed in a highly oxidizing environment (due to the presence of flowing oxygen and highly electropositive $Li^+$), and the reduction of $Co^{3+}$ to $Co^{2+}$ is unlikely. Another proposal is that the excess Li goes on the Co site and is accompanied by the formation of charge-compensating oxygen vacancies to maintain the $Co^{3+}$ state. In this case the magnetism is proposed to be due to a low-spin to high spin-transition on $Co^{3+}$ induced by the disorder [37]. There is no evidence, however, for the presence of oxygen vacancies in this compound. The most likely explanation is that the excess lithium ions substitute for cobalt in the stoichiometric $CoO_2$ layer, resulting in $Li_{1+x}Co_{1-x}O_2$ ($x>0$), and the introduction of magnetic $Co^{4+}$. A quantitative determination of the dependence of the magnetic moment on excess Li concentration in carefully synthesized and characterized Li-excess $LiCoO_2$ would be of further interest to resolve the actual microscopic mechanism of moment formation for this phase.

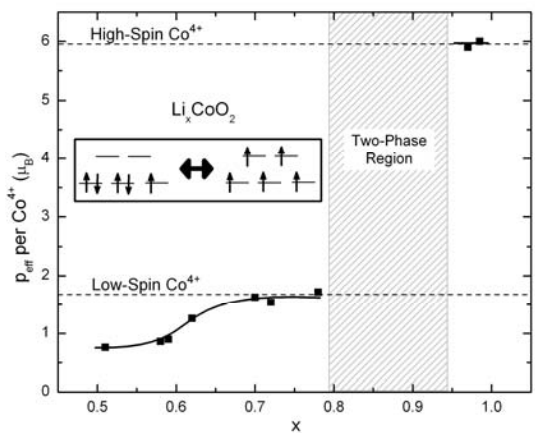

**Fig. 5**: The variation of the effective magnetic moment per $Co^{4+}$ ion with $x$ for $Li_xCoO_2$. A high spin to low spin transition is observed. Insert: schematic representation of the change in spin configuration.

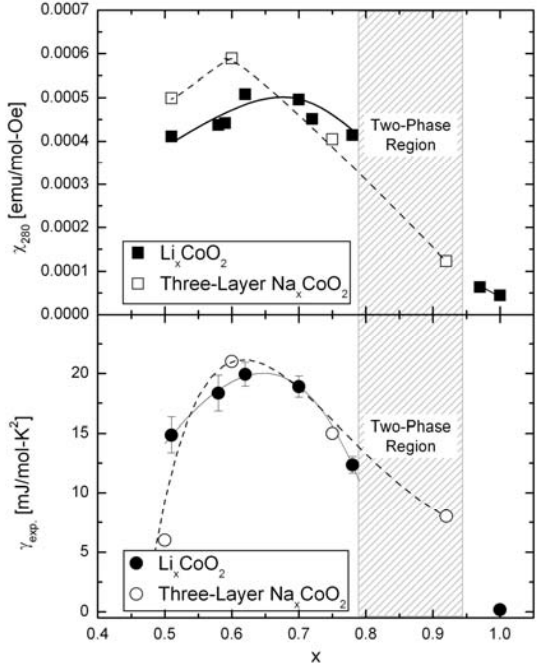

**Fig. 6**: Upper panel: The variation of the magnetic susceptibility at 280 K ($\chi_{280}$) with $x$ for both $Li_xCoO_2$ and three-layer $Na_xCoO_2$ [10]. Lower panel: The variation of the electronic contribution to the specific heat ($\gamma$) with $x$ for both $Li_xCoO_2$ and three-layer $Na_xCoO_2$ [10]. The shaded portion of this figure represents the two phase region in the $Li_xCoO_2$ system.

**Magnetic properties of $Li_xCoO_2$**

The magnetic properties of $Li_xCoO_2$ were studied for $x$=1.00, 0.97, 0.78, 0.72, 0.70, 0.62, 0.59, and 0.51. The temperature dependent susceptibility results are displayed in Figure 2. The data show several general features. Firstly, some of the samples display a small anomaly in the susceptibility at T = 175 K. Secondly, no magnetic ordering is observed in any of the samples down to the lowest temperature measured (T = 5 K); apart from the anomalies at 175 K, all the $Li_xCoO_2$ samples exhibit Curie-Weiss behavior. Finally, the temperature-independent magnetic contribution to the susceptibility increases considerably as the Li content in $Li_xCoO_2$ decreases to compositions below the two-phase region ($0.5 \leq x \leq 0.78$).

Several previous studies have reported the presence of a susceptibility anomaly at 175K in $Li_xCoO_2$ compounds [33-36]. Because this anomaly has been observed in samples with widely ranging Li contents, from $x$=0.95 [33] to $x$=0.50, and because the temperature of the anomaly is independent of the Li concentration over that very broad range of composition, we investigated whether the origin might be a magnetic impurity.

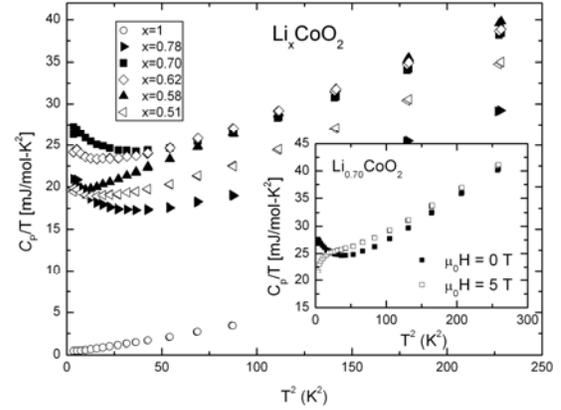

**Fig. 7**: The specific heat for $Li_xCoO_2$ plotted as $C_P/T$ versus $T^2$. Inset: The specific heat for $Li_{0.70}CoO_2$ with H = 0 T and H = 5 T.

The compounds CoO, $Co_3O_4$ and $Li_{0.1}Co_{0.9}O$ were tested, as was a mixed-phase sample at composition $Li_2CoO_x$, and none showed a 175 K anomaly. In addition, we re-measured the magnetic susceptibility after four months storage of three of the samples in our study ($x$=0.78, 0.72 and 0.51) that showed the magnetic anomaly. The data for one of the samples are shown in the inset to figure 2. Though the crystallographic cell parameters of the samples and overall behavior of the susceptibility were unchanged, indicating that the compounds had not decomposed or changed overall compositions substantially, the anomaly was strongly suppressed in all cases. This suggests that the anomaly may be due either to the presence of a small amount of an impurity phase that decomposes slowly in air, or that it is intrinsic but sensitive to a subtle change in the structure of the compound over time, as would happen for example if the ordering of the Li array was changing, or if the samples were becoming more chemically homogeneous with time due to Li diffusion between particles. Further work to determine the origin of this anomaly would be of interest.

Figure 2 shows that there is significant local moment character displayed in the susceptibilities of $Li_xCoO_2$ at low temperatures. The relationship between lithium content and the effective magnetic moment ($p_{eff}$) was determined by performing Curie-Weiss fits on the low temperature data (T≤75 K). The fitted $\chi_0$ values were used to plot the temperature dependence of $1/(\chi - \chi_0)$ for each of the samples. The results are shown in Figure 3 for the samples above and below the two-phase region. As expected for Curie-Weiss behavior, the plots are linear. The magnetic characteristics obtained from the fits are presented in Table 1.

Figure 4 shows the variation of the effective magnetic moment per formula unit ($p_{eff}$) with lithium content: $p_{eff}$ increases considerably as lithium is removed until a critical composition is reached ($x \approx 0.65$), then decreases upon further lithium extraction. As Table 1 and Figure 5 show, $\theta_W$ is small and negative for all of the $Li_xCoO_2$ samples, suggesting that the $Li_xCoO_2$ family in the range $0.5 \leq x \leq 1.0$ is characterized by weak antiferromagnetic interactions that become somewhat stronger as more lithium is extracted.

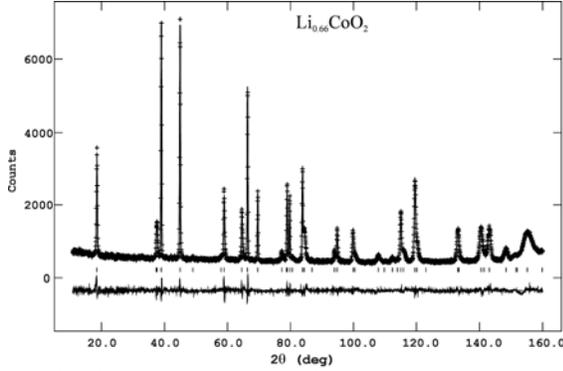

**Fig. 8.** Observed (points) and calculated (solid line) powder neutron diffraction pattern for $Li_{0.66}CoO_2$ at 298 K. Tic marks indicate the presence of a calculated peak. The lower curve shows the difference between observed and calculated intensities.

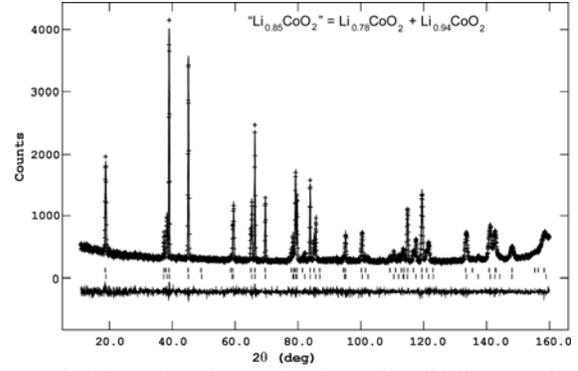

**Fig. 9.** Observed (points) and calculated (solid line) powder neutron diffraction pattern for $Li_{0.88}CoO_2$ at 298 K. This composition is in the two phase region. Tic marks indicate the presence of calculated peaks for the border compositions $Li_{0.94}CoO_2$ and $Li_{0.78}CoO_2$. The lower curve shows the difference between observed and calculated intensitites

The extraction of a lithium ion is accompanied by the formal oxidation of a cobalt atom from the non-magnetic +3 state to the magnetic +4 state, and so, if the unpaired spins remain completely localized, $p_{eff}$ is expected to increase continuously with decreasing lithium content in $Li_xCoO_2$. The fact that $p_{eff}$ *decreases* as $x$ decreases below 0.65 indicates that there is a changing ratio of local and non-local spins in $Li_xCoO_2$ as a function of composition This is demonstrated in Figure 5, which plots the effective magnetic moment per $Co^{4+}$ ion for $Li_xCoO_2$. If the unpaired spins are localized at all $x$, then the effective magnetic moment per $Co^{4+}$ should stay constant across the series. As Figure 5 shows, the effective magnetic moment per $Co^{4+}$ in the composition range $0.7 \leq x \leq 0.8$ is close to the expected low spin $Co^{4+}$ spin ½ value of 1.73 $\mu_B$ [53] and remains relatively constant as $x$ first decreases below 0.8. When $x$ drops below 0.65, however, the effective moment per $Co^{4+}$ decreases significantly. We attribute this drop in effective moment to a change in the localization of the unpaired spins at this composition.

As Table 1 shows, the nominally $Li_{0.97}CoO_2$ sample exhibits an effective magnetic moment of 0.23 $\mu_B$ per cobalt atom. For $x = 0.97$, with 3% of the cobalt atoms in the magnetic +4 oxidation state, this corresponds to a magnetic moment of 7.67 $\mu_B$ per $Co^{4+}$—a value much too large for a low-spin (spin ½) $d^5$ species. High-spin $Co^{4+}$ has five unpaired electrons and is expected to exhibit an effective magnetic moment of 5.92 $\mu_B$ : our observed effective moment per $Co^{4+}$ for $Li_{0.97}CoO_2$ is even larger. This difference is due to the precision to which the composition is known in the $Li_{0.97}CoO_2$ sample. If the sample instead has a composition of $x = 0.961$, within the error range of the determined composition of $x = 0.97 \pm 0.01$, then the effective magnetic moment per $Co^{4+}$ would be 5.9 $\mu_B$. Our results therefore indicate the presence of high-spin $Co^{4+}$ in $Li_xCoO_2$ for this composition. (In order for a low-spin $Li_xCoO_2$ species to exhibit an effective magnetic moment of 0.23 $\mu_B$, the fraction of $Co^{4+}$ needed would correspond to a Li composition of $Li_{0.868}CoO_2$, well outside the error range of the composition for the sample, and in the chemical two-phase region, which is not the case.) Thus, the nominally $Li_{0.97}CoO_2$ sample contains high-spin $Co^{4+}$ and has an actual composition of $Li_{0.96}CoO_2$.

Similarly, the nominally stoichiometric $LiCoO_2$ sample also exhibits a small magnetic moment, suggesting that it is not perfectly stoichiometric and contains a small number of $Co^{4+}$ ions. In order for a low-spin $Li_xCoO_2$ species to exhibit the observed effective magnetic moment of 0.09 $\mu_B$, the fraction of $Co^{4+}$ present would correspond to a lithium content of $x = 0.948$, a value again well outside the error range of the composition for the sample ($x = 1.00 \pm 0.02$). If the $Co^{4+}$ ions are high-spin, however, then the fraction of $Co^{4+}$ is 0.015. This corresponds to a composition of $Li_{0.985}CoO_2$, or $Li_{1.008}Co_{0.992}O_2$ if excess Li is accommodated, both of which are within the error of the observed composition. Thus, our results indicate that the high $x$ compositions in $Li_xCoO_2$ contain high-spin $Co^{+4}$. This analysis implies that the magnetic $Co^{4+}$ introduced by using excess Li in the synthesis of LiCoO$_2$, illustrated in Figure 1, is also high spin.

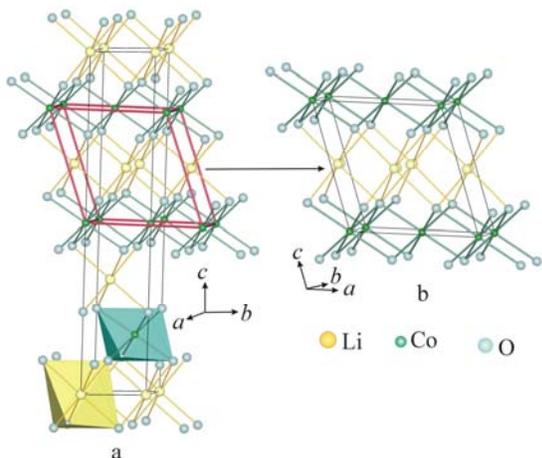

**Fig. 10.** (a) The rhombohedral crystal structure of LiCoO$_2$. (b) The monoclinic crystal structure of Li$_{0.51}$CoO$_2$. The partially occupied Li positions, slightly displaced from the centers of the LiO$_6$ octahedra, are shown. The relationship between the three layer rhombohedral unit cell and the monoclinic unit cell is also shown.

Analysis of the magnetic data, therefore, indicates that Li$_x$CoO$_2$ samples for low $x$ are low-spin, while samples at high $x$ are high-spin. This high-spin to low-spin transition, which occurs at the chemical phase boundary, is illustrated in Figure 5. The nature of the spin state of Co in oxides, especially in the Na$_x$CoO$_2$ and Li$_x$CoO$_2$ systems, has been the subject of some debate. Numerous recent studies have argued that both Co$^{3+}$ and Co$^{4+}$ are low-spin in the whole Na$_x$CoO$_2$ system [39,54-56] and Li$_x$CoO$_2$ system [35], while some have claimed to observe intermediate spin [37] or high spin [34] Co. The case has been argued based on the relative stabilities of electron configurations for Co in octahedral coordination that high-spin Co$^{4+}$ should be stable in these compounds [57]. The fact that Co$^{4+}$ in Li$_x$CoO$_2$ above the two-phase region is high spin suggests that these arguments have merit. The structural data, described in a later section, suggest that the spin state transition in Li$_x$CoO$_2$ could be driven by a volume effect, since the low spin Co$^{4+}$ is found in the part of system with shorter Co-O bondlengths, but the differences in bondlengths are subtle and continuous across the transition and so the case for that origin for the spin state transition is weak. The current data appear to present the cleanest case for a high-spin to low-spin transition in the triangular layered CoO$_2$ family.

We employ the magnetic susceptibility at 280 K ($\chi_{280}$) as a measure of the temperature-independent paramagnetism arising from the conduction electrons. The upper panel of Figure 6 shows the variation of $\chi_{280}$ with lithium content. Though $\chi_{280}$ does not vary dramatically among the highly delithiated compounds ($0.51 \leq x \leq 0.78$), it displays a weak peak at a composition $x \approx 0.65$. The temperature-independent susceptibility is an order of magnitude lower for the compounds with Li contents greater than 0.95, indicating that the high $x$ materials have very few free electrons. The upper panel of Figure 6 also shows the variation of $\chi_{280}$ for three-layer Na$_x$CoO$_2$ [10]. The chemical phase diagrams are different in the two cases, but similarities between the Li$_x$CoO$_2$ and Na$_x$CoO$_2$ systems are seen. The absolute values of $\chi_{280}$ in the range $0.5 \leq x \leq 1.00$ are quite similar, and the temperature-independent susceptibility increases substantially when more than 10% of the alkali metal is removed from the stoichiometric Co$^{3+}$ compounds in both systems. The temperature-independent part of the susceptibility is maximized at a similar composition in both systems ($x \approx 0.65$ for Li$_x$CoO$_2$, and $x \approx 0.60$ for three-layer Na$_x$CoO$_2$), though in Na$_x$CoO$_2$ it is more strongly peaked.

Figure 7 displays the low temperature specific heat data for six Li$_x$CoO$_2$ samples, plotted as $C_p/T$ vs. $T^2$. The temperature dependence of the specific heat in systems where there are electronic and lattice contributions to the specific heat at low temperatures can be written as: $C_p/T = \gamma + \beta T^2$, where $\beta$ and $\gamma$ are parameters describing the size of the lattice and electronic contributions to the specific heat, respectively. The values of γ obtained from linear fits to C/T vs. $T^2$ plots are presented in Table 1. The data in Figure 7 are linear at higher temperatures ($T^2$ > 75 K$^2$), while at lower temperatures, significant deviations from linearity occur, suggesting that magnetic fluctuations contribute significantly to the specific heat at low temperatures. To further investigate this, specific heat was measured under a 5 T applied field for $x$ = 0.70 (Figure 7). The applied field changes the behavior of the specific heat at low temperatures, suggesting that the deviation from linearity is indeed due to magnetic fluctuations. Further study of the low temperature properties of these compounds may therefore be of interest.

**Table 2**: Refined unit cell parameters for Li$_x$CoO$_2$ from the neutron diffraction data.

| Compound | Space Group | Cell Constants (Å) | Unit Cell Volume (Å$^3$) | Volume/F.U. (Å$^3$) |
|---|---|---|---|---|
| LiCoO$_2$ | $R$-$3m$ (No. 166) | a = 2.81698(3)<br>c = 14.0646(1) | 96.655(2) | 32.22 |
| Li$_{0.94}$CoO$_2$ | $R$-$3m$ (No. 166) | a = 2.81593(6)<br>c = 14.0604(4) | 96.554(5) | 32.18 |
| Li$_{0.78}$CoO$_2$ | $R$-$3m$ (No. 166) | a = 2.81496(9)<br>c = 14.1609(5) | 97.178(6) | 32.39 |
| Li$_{0.75}$CoO$_2$ | $R$-$3m$ (No. 166) | a = 2.81185(3)<br>c = 14.2264(4) | 97.412(3) | 32.47 |
| Li$_{0.66}$CoO$_2$ | $R$-$3m$ (No. 166) | a = 2.81172(3)<br>c = 14.2863(4) | 97.812(3) | 32.60 |
| Li$_{0.51}$CoO$_2$ | $C$ $2/m$ (No. 12) | a = 4.8645(1)<br>b = 2.80964(7)<br>c = 5.0551(1)<br>β = 107.908(1)° | 65.744(4) | 32.87 |

The lower panel of Figure 6 depicts the variation of γ with lithium content in Li$_x$CoO$_2$ and demonstrates a similarity between the behavior of $\chi_{280}$ and γ with composition. The electronic contribution to the specific heat (γ) increases continuously with decreasing lithium content until γ is maximized at the previously observed critical composition ($x \approx 0.65$). Comparison to the data for Na$_x$CoO$_2$ shows generally similar behavior in the two systems.

**Crystal Structures of Li$_x$CoO$_2$ and Comparison to Na$_x$CoO$_2$**

The refinements of the structures of Li$_x$CoO$_2$ from the neutron powder diffraction data were straightforward. The compositions of the samples were fixed at the analytically determined chemical compositions. The structure of LiCoO$_2$, with edge sharing layers of CoO$_6$ octahedra interleaved with layers of edge sharing LiO$_6$ octahedra, provided the initial model for the fits. The refined crystallographic cell parameters from the neutron powder diffraction data are presented in Table 2. Structures for Li$_x$CoO$_2$ for $x$= 1.0, 0.94, 0.78, 0.75, and 0.66 were very well described by the rhombohedral three-layer LiCoO$_2$ structure in space group $R$-$3m$ (#166). A two phase refinement at overall composition Li$_{0.85}$CoO$_2$ was employed to determine the structures of the border compositions of the two phase region, $x$=0.94 and $x$=0.78. The symmetry of Li$_{0.51}$CoO$_2$ is monoclinic, space group $C$ 2/$m$ (#12).

The published model for LiCoO$_2$ resulted in an excellent fit to our data for compositions on the high $x$ side of the two-phase region, $x$=1.0 and 0.94. However, refinements for lower Li content phases showed unacceptably large thermal vibration parameters for the Li atoms, with thermal ellipsoids much larger in the basal plane than perpendicular to the plane. This indicates that the Li are displaced in-

**Table 3a** Structural parameters for trigonal Li$_x$CoO$_2$ ($x$=1, 0.85, 0.75, and 0.66) at 295 K. Space group: $R\bar{3}m$. Atomic positions: Li: 3$a$(0 0 0), or 18$f$ ($x$ 0 0) Co: 3$b$ (0,0,1/2), and O: 6$c$ (0 0 $z$).

| Atom | Parameters | ($x$= 1.0 | 0.94, [63.2(2)%]* | 0.78, [36.8(2)%]* | 0.75 | 0.66 |
|---|---|---|---|---|---|---|
| | $a$ (Å) | 2.81698(3) | 2.81593(6) | 2.81496(9) | 2.81185(3) | 2.81172(3) |
| | $c$ (Å) | 14.0646(1) | 14.0604(4) | 14.1609(5) | 14.2264(4) | 14.2863(4) |
| | $V$ (Å$^3$) | 96.655(2) | 96.554(5) | 97.178(6) | 97.412(3) | 97.812(3) |
| Li | $x$ | | | 0.05(1) | 0.066(5) | 0.077(4) |
| | $U_{iso}$ (Å$^2$) | 0.0158(3) | 0.013(1) | | 0.017(2) | 0.017(3) |
| Co | $U_{11}$, $U_{22}$ (Å$^2$) | 0.0074(3) | 0.0062(5) | 0.0102(6) | 0.0108(5) | |
| | $U_{33}$ (Å$^2$) | 0.0075(5) | 0.014(1) | | 0.012(1) | 0.012(1) |
| | $U_{12}$ (Å$^2$) | 0.0037(2) | 0.0031(2) | 0.0051(3) | 0.0054(3) | |
| O | $z$ | 0.23949(3) | 0.23942(7) | 0.2380(1) | 0.23685(6) | 0.23620(6) |
| | $U_{11}$, $U_{22}$ (Å$^2$) | 0.0087(1) | 0.0082(5) | 0.0114(3) | 0.0114(2) | |
| | $U_{33}$ (Å$^2$) | 0.0102(2) | 0.0124(4) | 0.0135(5) | 0.0156(5) | |
| | $U_{12}$ (Å$^2$) | 0.0044(1) | 0.0041(4) | 0.0057(1) | 0.0057(1) | |
| | $R_p$ (%) | 3.66 | 3.81 | | 4.76 | 4.05 |
| | $wR_p$ (%) | 4.54 | 4.73 | | 6.64 | 5.12 |
| | $\chi^2$ | 1.054 | 0.893 | | 1.769 | 1.794 |

* Two phase sample. The temperature factors for Li, Co, and O in two phases were constrained to be equal.

**Table 3b.** Structural parameters for monoclinic Li$_{0.51}$CoO$_2$ at 295 K. Space group: $C2/m$. $a$=4.8645(1) Å, $b$=2.80964(7) Å, $c$=5.0551(1) Å, $\beta$=107.908(1)°, $V$=64.744(4) Å$^3$.

| Atom | Site | $x$ | $y$ | $z$ | $n$ | [$U_{iso}$] $U_{11}$(Å$^2$) | $U_{22}$(Å$^2$) | $U_{33}$(Å$^2$) | $U_{13}$(Å$^2$) |
|---|---|---|---|---|---|---|---|---|---|
| Li | 4h | 0 | 0.582(3) | ½ | 0.255 | [0.018(3)] | | | |
| Co | 2a | 0 | 0 | 0 | 1 | 0.006(1) | 0.011(1) | 0.019(2) | 0.004(1) |
| O | 4i | 0.7312(2) | 0 | 0.2042(2) | 1 | 0.0093(6) | 0.012(5) | 0.0197(6) | 0.0058(4) |

$R_p$ (%) 3.32
$wR_p$ (%) 4.35
$\chi^2$ 1.842

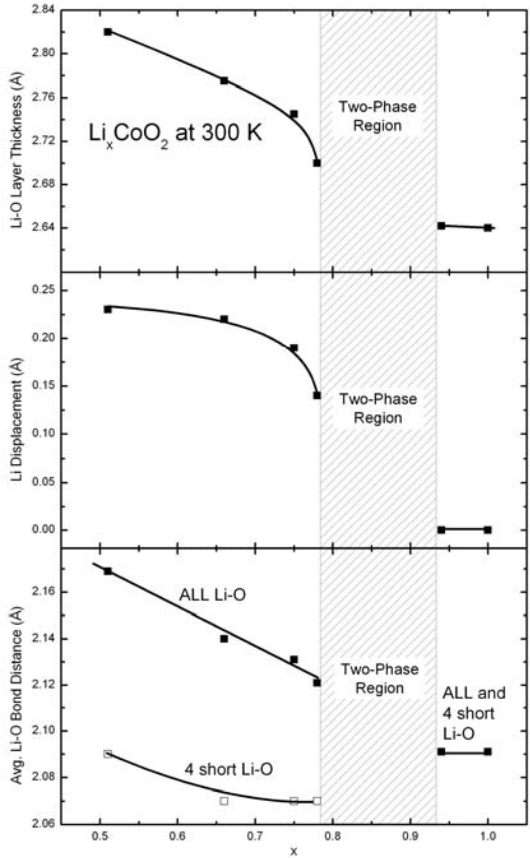

**Fig. 11** The Li-O plane characteristics in $Li_xCoO_2$. Upper Panel: Variation of the Li layer thickness with *x*. Middle panel: The magnitude of the displacement of the lithium ions from their ideal octahedral sites. Lower panel: The average Li-O bond distance taken over all oxygens, and the average taken over four nearest oxygens.

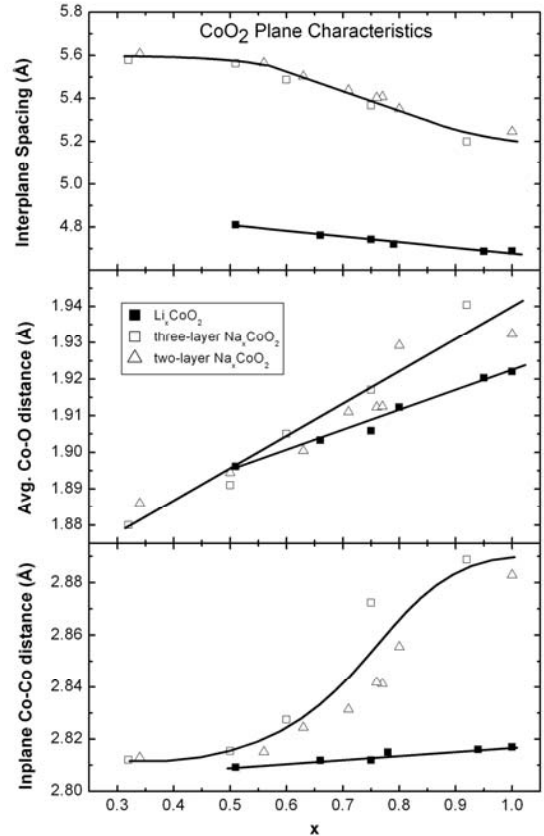

**12:** Comparison of the $CoO_2$ plane characteristics in $Li_xCoO_2$ and $Na_xCoO_2$ [4,10]. Upper panel: the distance between $CoO_2$ layers as a function of Li (Na) content. Middle panel: the variation of average Co-O bond length with Li(Na) content. Lower panel: The variation in in-plane Co-Co distance as a function of Li (Na) content.

plane from the centers of their ideal octahedra: the (0,0,0) site for the rhombohedral phases and the (0, ½, ½) site for monoclinic $Li_{0.51}CoO_2$. The refinement of this displacement was significant in all cases and the resulting thermal vibration parameters for the Li assumed normal values. Displacements of the alkali ions from the centers of the coordination polyhedra are commonly seen in $Na_xCoO_2$ [4] and have been seen in a different polymorph of $Li_xCoO_2$ [15]. The agreements between the refined models and the data are excellent (see Figures 8 and 9). The refined atomic positions for lithium, cobalt, and oxygen for all compounds are presented in Tables 3a and 3b, and selected bond lengths and bond angles are displayed in Table 4.

Monoclinic symmetry crystallographic cells are commonly encountered in layered crystals structures of this type. This occurs when slight shifts of adjacent $CoO_2$ layers from the ideal positions in a three-layer rhombohedral cell disrupt the three fold symmetry. The $Na_xCoO_2$ system shows this behavior, for example [10]. For the $Li_xCoO_2$ system, this type of monoclinic cell has been previously reported for $0.46 \leq x \leq 0.54$. An electron diffraction study of this phase [28] indicated a monoclinic cell of *P2/m* symmetry that was modeled with lithium ions occupying half of the available octahedral sites between the $CoO_2$ layers in an ordered fashion. In our powder neutron diffraction patterns of $Li_{0.51}CoO_2$, however, the extra reflections that would result from the presence of ordered lithium ions were not observed. This suggests that if present the Li ordering is only over short ranges and can only be observed by electron diffraction. The refinements were therefore carried out in the *C2/m* space group (#12), in which the Li fills half the available octahedral sites between the $CoO_2$ layers in a disordered fashion. The relationship between the rhombohedral and monoclinic structures for $Li_xCoO_2$ is shown in Figure 10.

The structural characteristics of the Li-O plane in $Li_xCoO_2$ are presented in Figure 11. The upper panel shows the composition dependent variation of the Li-O layer thickness, determined by calculating the perpendicular distance between adjacent planes of oxygen anions separated by Li. As is seen in other systems, the removal of the strongly bonding alkali ions from between the $CoO_2$ layers causes the layers to move further apart as their repulsion overcomes the bonding forces holding them together. The $LiO_6$ coordination polyhedra, which are already large in $LiCoO_2$ (Li-O bond length 2.09 A), further increase in size as the layers move apart. This increasing size of the interplane spacing drives the Li off-center in their coordination polyhedra so they can preserve favorable Li-O bonding distances. With a significant fraction of the Li sites vacant in the composition region $0.51 < x < 0.78$ this is possible because in-plane repulsion of neighboring Li is relaxed. The middle panel of Figure 11 plots the magnitude of the displacement of the lithium ions from their ideal octahedral sites versus lithium content; they become increasingly displaced as the Li-O layer expands. As the lower panel of Figure 11 shows, the average Li-O bond distance increases substantially as lithium is removed. That is, the $LiO_6$ octahedra become stretched significantly in the $c$-direction as $x$ decreases, resulting in longer average Li-O bonds. However, the displacements are such that the Li stays close to four of its six neighboring oxygens in spite of the expansion of the oxygen polyhedron, maximizing the bonding interactions with those four oxygens at the expense of the interactions with the other two. Although the average Li-O bond distance increases substantially as $x$ decreases, the average distance from the Li to the four nearest oxygens remains relatively constant throughout the series, seen in the lower panel to Figure 11. Thus, once the lithium layer has been partially depopulated, the lithium ions displace from their ideal octahedral sites to maintain strong bonding interactions with some of their oxygen neighbors.

**Table 4**: Selected Bond Lengths and Bond Angles for $Li_xCoO_2$

| Compound | Co-O bond distance (Å) | O-Co-O bond angle (deg.) | Li-O bond distance (Å) | In-plane Co-Co distance (Å) |
|---|---|---|---|---|
| $LiCoO_2$ | 1.9220(2) ×6 | 94.24(1) ×6<br>85.75(1) ×6 | 2.0946(3) ×6 | 2.81698(3) |
| $Li_{0.94}CoO_2$ | 1.9210(5) ×6 | 94.27(3) ×6<br>85.73(3) ×6 | 2.0944(6) ×6 | 2.81593(6) |
| $Li_{0.78}CoO_2$ | 1.9132(8) ×6 | 94.72(5) ×6<br>85.28(5) ×6 | 2.21(2) ×2<br>2.12(2) ×2<br>2.02(2) ×2 | 2.81496(9) |
| $Li_{0.75}CoO_2$ | 1.9059(5) ×6 | 95.07(3) ×6<br>84.93(3) ×6 | 2.25(1) ×2<br>2.134(1) ×2<br>2.007(9) ×2 | 2.81185(3) |
| $Li_{0.66}CoO_2$ | 1.9032(5) ×6 | 95.24(3) ×6<br>84.76(3) ×6 | 2.284(8) ×2<br>2.147(1) ×2<br>2.000(7) ×2 | 2.81172(3) |
| $Li_{0.51}CoO_2$ | 1.8986(9) ×2<br>1.8948(6) ×4<br>Avg: 1.896 | 95.54(4) ×4<br>95.70(4) ×2<br>84.46(4) ×4<br>84.30(4) ×2 | 2.328(7) ×2<br>2.142(1) ×2<br>2.030(5) ×2 | 2.80909(7) |

**Table 5**: Summary of significant structural parameters in $Li_xCoO_2$.

| Lithium Content ($x$) | Li layer thickness (Å) | Avg. Li-O distance (Å) | Li displacement (Å) | Avg. 4 Shortest Li-O Bonds (Å) | Observed $CoO_2$ layer thickness $T_{obs}$ (Å) | Ideal $CoO_2$ layer thickness $T_{ideal}$ (Å) | $T_{obs}/T_{ideal}$ |
|---|---|---|---|---|---|---|---|
| 1.00 | 2.640 | 2.09 | 0 | 2.09 | 2.048 | 2.300 | 0.890 |
| 0.94 | 2.641 | 2.09 | 0 | 2.09 | 2.046 | 2.299 | 0.890 |
| 0.78 | 2.700 | 2.12 | 0.14(3) | 2.07 | 2.020 | 2.298 | 0.879 |
| 0.75 | 2.745 | 2.13 | 0.19(1) | 2.07 | 1.997 | 2.296 | 0.870 |
| 0.66 | 2.775 | 2.14 | 0.22(1) | 2.07 | 1.987 | 2.296 | 0.865 |
| 0.51 | 2.819 | 2.17 | 0.23(1) | 2.09 | 1.965 | 2.294 | 0.857 |

Neutron diffraction studies of the two- and three-layer $Na_xCoO_2$ systems have quantified significant structural trends in the $CoO_2$ layers as a function of alkali ion content, specifically changes in Co-O bond length, in-plane Co-Co distance, and the interplane spacing [4,10]. The variations of these characteristics for $Li_xCoO_2$ are compared with those in two- and three-layer $Na_xCoO_2$ in Figure 12. These three characteristics change much more dramatically with alkali ion content in $Na_xCoO_2$ than in $Li_xCoO_2$. On increasing $x$ from 0.5 to 1.0, the Co-O bond distance increases by 0.45 Å for $Na_xCoO_2$, but only by 0.26 Å for $Li_xCoO_2$. Similarly, the Co-Co distance within the $CoO_2$ plane, reflecting the in-plane size of the $CoO_6$ octahedra, increases by 0.74 Å for $Na_xCoO_2$, but only by 0.08 Å for $Li_xCoO_2$ - a nearly ten-fold difference. Finally, the change in spacing between $CoO_2$ planes is also significantly different, decreasing by 3.7 Å for $Na_xCoO_2$ and by half that, 1.8 Å, for $Li_xCoO_2$.

Near $x=0.5$, the Co-O bond lengths and in-plane Co-Co distances are quite similar in the sodium and lithium systems (Figure 12). As described above, these two structural parameters increase substantially with increasing $x$ in $Na_xCoO_2$, while only changing slightly in $Li_xCoO_2$. Most dramatically, the $CoO_2$ layer expands considerably in the in-plane dimension in the sodium system, but not in the lithium system. It may be that the relatively large size of the sodium ions (radius = 1.04 Å [58]) causes crowding that gives rise to the dramatic expansion of the $CoO_2$ layer in $Na_xCoO_2$: as the sodium layer becomes filled, the $CoO_2$ layer must stretch in-plane to accommodate the coordination of bulky sodium ions to the oxygen anions. Alternatively it could be that rather than the sodium ions being too large, the lithium ions are too small: the more compact lithium ions (radius = 0.74 Å [58]) may restrict what the electronic energies would prefer, the in-plane expansion of the $CoO_2$ layer, because this expansion would overstretch the Li-O bonds. Whatever the underlying cause, the geometries of the $CoO_2$ layers are very different for $0.5<x<1.0$ in $Li_xCoO_2$ and $Na_xCoO_2$.

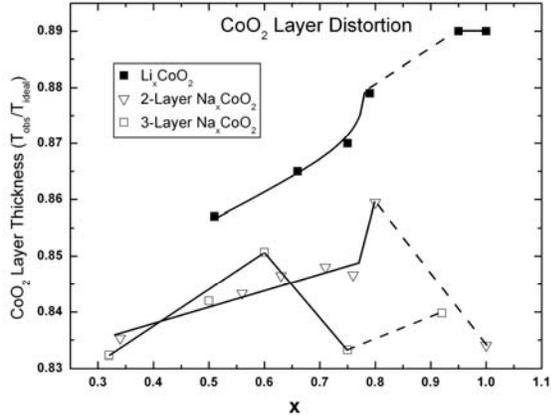

**Fig. 13**: The variation of the $T_{obs}/T_{ideal}$ ratio with $x$ for $Li_xCoO_2$ and two- and three-layer $Na_xCoO_2$ [4,10].

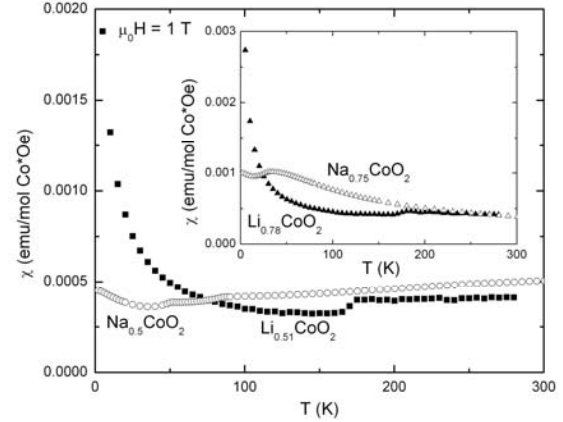

**Fig. 14:** Comparison of the temperature dependent magnetic susceptibilities for selected compositions of $Na_xCoO_2$ and $Li_xCoO_2$.

**Discussion**

The structural characteristics of the $CoO_2$ layers in $Na_xCoO_2$ and $Li_xCoO_2$ directly impact their electronic systems. The in-plane expansion of the $CoO_2$ layer observed when x>0.75 in $Na_xCoO_2$ (Figure 12) suggests that significant changes in electronic band occupancy occur at that composition, whereas the much smaller change in size of the $CoO_2$ plane in $Li_xCoO_2$ suggests that the electronic system does not undergo qualitative changes in that case. We have previously proposed that the distortion of the $CoO_6$ octahedron is an important indicator of the electronic state of $Na_xCoO_2$ [10], due to the expectation that the shape of the $CoO_6$ octahedra will impact the relative energies of Co $t_{2g}$ and $e_g$ suborbitals. To quantify this distortion, the in-plane Co-Co distance can be used to calculate the $CoO_2$ layer thickness that would be expected if the $CoO_6$ octahedra were ideal in shape – a situation that would lead to degeneracy among the three $t_{2g}$ orbitals. This ideal layer thickness is then compared to the actual observed thickness of the $CoO_2$ layer to quantify how much the $t_{2g}$ degeneracies might be lifted by the structural distortion. The relationship between the in-plane Co-Co distance ($d_{Co-Co}$) and the thickness of a layer of ideal edge-sharing $CoO_6$ octahedra ($T_{ideal}$) is:

$T_{ideal} = (\sqrt{6}/3) \, d_{Co-Co}$. By determining the ratio of the observed layer thickness to the ideal layer thickness $T_{obs}/T_{ideal}$, as enumerated in Table 5, the degree of the distortion of the $CoO_6$ octahedra can be obtained. If this ratio is less than 1, as it is in these systems, then the $CoO_6$ octahedra are compressed perpendicular to the $CoO_2$ planes.

Comparison of the variation of $T_{obs}/T_{ideal}$ with alkali content for three-layer $Li_xCoO_2$, two-layer $Na_xCoO_2$, and three-layer $Na_xCoO_2$ is shown in Figure 13. As lithium content decreases in $Li_xCoO_2$, the $CoO_2$ octahedra are monotonically more compressed perpendicular to the plane. The behavior in the $Na_xCoO_2$ systems is much more complex, suggesting significant electronic reconfigurations across the series. Further, the $CoO_6$ octahedra in $Li_xCoO_2$ are more regular: the layers are 85-90% of their ideal thicknesses, whereas those in the $Na_xCoO_2$ systems are 83-85% of their ideal thicknesses. Thus, at least in the range $0.5 \leq x \leq 1.0$, the $CoO_6$ octahedra are significantly more distorted in $Na_xCoO_2$ than in $Li_xCoO_2$, suggesting that the electronic systems are quite different in these families.

The magnetic behavior of the $Li_xCoO_2$ compounds, shown in Figure 2, differs significantly from the reported magnetic behavior of $Na_xCoO_2$. Figure 14 compares the temperature dependence of the magnetic susceptibilities for several $Li_xCoO_2$ and $Na_xCoO_2$ compounds synthesized in this laboratory [4,10]. The $Na_xCoO_2$ family displays more unusual magnetic behavior. $Na_{0.5}CoO_2$, for instance, departs significantly from Curie-Weiss like behavior, whereas $Li_{0.51}CoO_2$ shows ordinary paramagnetic behavior. Three-layer $Na_{0.78}CoO_2$ exhibits a magnetic ordering transition near T = 35 K, as do two-layer $Na_xCoO_2$ materials near this composition (insert). These magnetic transitions are not observed in $Li_xCoO_2$, though some kind of anomaly occurs at 175K across the series. $Li_xCoO_2$ appears to display conventional magnetic behavior, while the behavior of $Na_xCoO_2$ indicates that more unconventional electronic factors are in play. The structural rigidity of $Li_xCoO_2$, yielding a more geometrically ideal $CoO_2$ plane whose characteristics do not vary substantially across the series, could be the underlying cause account for the relatively simple magnetic behavior of that system.

**Conclusions**

A study of the magnetic properties and structural characteristics of $Li_xCoO_2$ has been reported. We postulate that the structural variety in the $Na_xCoO_2$ family gives rise to the many exotic magnetic and electronic properties that have made it the subject of intense study, and that the lack of this structural variety in $Li_xCoO_2$ accounts for its more conventional properties. Analysis of the magnetic data shows that the spin state of $Co^{4+}$ in $Li_xCoO_2$ changes from high-spin at compositions above the two-phase region to low-spin at compositions below the two-phase region. It is unclear why this high-spin to low-spin transition is not observed in the high $x$ region of $Na_xCoO_2$, since the larger Co-O octahedra observed there would be expected to be more favorable to the existence of high-spin $Co^{4+}$. The lack of high-spin behavior in $Na_xCoO_2$ in the range $0.94 \leq x \leq 1.00$ must be related to other features of the local environment of $Co^{4+}$ in the sodium system, such as the significant compression of $CoO_6$ octahedra, which is not observed in $Li_xCoO_2$. Comparative study of $Li_xCoO2$ and $Na_xCoO_2$ by methods sensitive to the Co spin state would be of significant future interest. In all $A_xCoO_2$ systems studied thus far, a chemical two-phase region is observed for x ≈ 0.8-0.9. It will be of interest to determine whether this phase separation is chemically or electronically driven. Finally, though this study has established correlations between the structures and properties of $Na_xCoO_2$ and $Li_xCoO_2$, it has not elucidated the basic physics that underlies those connections. Because the triangular $CoO_2$ planes found in these cobaltates differ substantially in phenomenology from the square planes commonly studied in Perovskites, further theoretical and experimental study of their structure-property correlations would clearly be of future interest.

**Acknowledgements**
The research at Princeton University is supported by the US Department of Energy, Division of Basic Energy Sciences, grant DOE-FG98-02-45706.